\documentclass[preprint,preprintnumbers,amsmath,amssymb]{revtex4}

\pdfoutput=1

\usepackage{graphicx}
\usepackage{dcolumn}
\usepackage{bm}

\newcommand{\clearfig}{\clearpage}

\newcommand{\speed}{\ensuremath{v_{\rm F}}}

\newcommand{\gate}{\ensuremath{V_{\rm G}}}
\newcommand{\bohr}{\ensuremath{\mu_{\rm B}}}
\newcommand{\gfactor}{\ensuremath{g}}         
\newcommand{\dos}{\ensuremath{ \frac{{\rm d}n}{{\rm d}\epsilon} }}
\newcommand{\voff}{\ensuremath{ \gate^{\rm offset} }}
\newcommand{\cond}{\ensuremath{ G }}

\newcommand{\corr}{\ensuremath{C_{[\delta \cond]}}}

\begin{document}

\title{Spin-resolved Quantum Interference in Graphene}

\author{Mark B. Lundeberg}
\thanks{to whom correspondence should be addressed: {\it mbl@physics.ubc.ca}}
\author{Joshua A. Folk}
\affiliation{Department of Physics and Astronomy, University of British Columbia, Vancouver, BC V6T
1Z4, Canada}


\date{\today}

\maketitle

{\bf
The unusual electronic properties of single-layer graphene\cite{novoselov} make it a promising material system for fundamental advances in physics, and an attractive platform for new device technologies. Graphene's spin transport properties are expected to be particularly interesting, with predictions for extremely long coherence times and intrinsic spin-polarized states at zero field\cite{yazyev-2007,PhysRevLett.95.226801,PhysRevB.72.174406,wimmer:177207}.
In order to test such predictions, it is necessary to measure the spin polarization of electrical currents in graphene.
Here, we resolve spin transport directly from conductance features that are caused by quantum interference. These features split visibly in an in-plane magnetic field, similar to Zeeman splitting in atomic and quantum dot systems\cite{zeeman,PhysRevLett.81.681}.
The spin-polarized conductance features that are the subject of this work may, in the future, lead to the development of graphene devices incorporating interference-based spin filters.

}

Spin currents have recently been measured in graphene using ferromagnetic contacts, an advance that shows promise for room-temperature graphene spintronics\cite{ferrographene,cho:123105,2007JaJAP..46L.605O}.  Unfortunately, 
many of the other tools that have been developed for studying spin-dependent phenomena in solid-state nanostructures (optical excitation, spin-resolved Coulomb blockade, etc.)
have proven difficult to implement in graphene.  The Zeeman-split conductance fluctuations that are the subject of this paper provide a way to detect spin-polarized currents with a straightforward electrical measurement.  This is the first report, to our knowledge, of visible spin-splitting of conductance fluctuations in any material.

Flakes of graphene were deposited using well-established mechanical exfoliation techniques\cite{novoselov} onto silicon wafers with a $\sim$300nm surface oxide, then contacted with Cr/Au leads (Fig.~1a).  The flake used for this experiment was selected optically and confirmed to be single layer by quantum Hall measurements (see supplement) showing the characteristic (n+1/2) Landau quantization\cite{novoselov}.  The two-terminal conductance measurements described below were performed in a dilution refrigerator with a base temperature of 20mK. The distance between the contacts was 3$\mu$m; the phase coherence length was estimated to equal or exceed sample dimensions. A two-axis magnet provided fields up to $B_\parallel$=12T in the plane of the graphene and fine control of the out-of-plane component up to $B_\perp$=120mT (Fig.~1a).

Carrier density $n$ was controlled capacitively with a voltage $\gate$ on the backgate, inducing a density $n=\alpha\gate$ in the graphene sheet. The proportionality constant, $\alpha = 8.05\pm 0.05 \times 10^{10}$ cm$^{-2}$/V, was extracted from the quantum Hall data. A Dirac resistance peak was observed when $\gate$ tuned the carrier density to the charge neutrality point at $V_0$$\approx$1V.  The conductance, $G$, was calculated after taking into account an estimated contact resistance of 3.2k$\Omega$. Conductance fluctuations, $\delta G$, were visible superimposed on the Dirac lineshape (Fig.~1b), and were extracted by subtracting an ensemble-averaged background (see supplement).

Graphene, like any metallic or semiconducting system, exhibits `universal' conductance fluctuations when transport is phase coherent\cite{beenakker,ucfg2,wlg4}.
These aperiodic fluctuations arise from quantum interference between all possible paths that charge carriers take as they traverse the device. The interference pattern is randomized by small changes in the relative phase between trajectories, which depends on the Aharonov-Bohm magnetic flux through the sample as well as on the Fermi wavevector\cite{beenakker}.  The conductance includes contributions from spin-up and spin-down carriers: $\cond = \cond_\uparrow+  \cond_\downarrow$.   In the absence of spin-orbit and many-body interactions, $\cond_{\uparrow/\downarrow}$ depends on the spin only through the Fermi wavevector, and therefore the density, of the respective carrier population.

Spin-up and spin-down carriers have the same density at zero field,
$n_\uparrow$=$n_\downarrow$. In a simple picture, then, zero-field conductance fluctuations include identical contributions from both populations (Fig.~1c). Applying a magnetic field partially polarizes the graphene -- it induces a difference between spin-up and spin-down carrier densities.   The total density $n=n_\uparrow$$+$$n_\downarrow=\alpha\gate$ is set by the backgate voltage; the difference in the densities $n_\uparrow$$-$$n_\downarrow = \frac{1}{2} \dos\gfactor\bohr B$ is set by the magnetic field, where $\dos$ is the density of states and $\gfactor\bohr B$ is the Zeeman energy.  Together these yield
\begin{equation}
 n_{\uparrow/\downarrow}=     \frac{\alpha}{2} \left( \gate \pm \frac{\voff}{2} \right),\\
~~~\voff = \frac{1}{\alpha}\frac{{\rm d}n}{{\rm d}\epsilon} \gfactor \bohr B
\end{equation}
As a result, spin-up and spin-down conductance contributions at finite field are offset in gate voltage by $\voff$ (Fig. 1d), leading to Zeeman splitting of interference features in a gate voltage trace $\cond(\gate)$ (Fig. 1e).

Zeeman splitting of graphene conductance fluctuations can be seen in Figs.~1F~and~1G.  The `V' shapes in the data correspond to spin-resolved conductance: left-moving (right-moving) arms reflect interference for a particular density of spin-down (spin-up) carriers.
 Although conductance fluctuations have been observed for several decades in a wide variety of materials\cite{wlg4,beenakker},
 a direct measurement of their Zeeman splitting has never before been reported.
We speculate that this effect is visible in graphene due in part to its atomic-scale flatness and low density of states.

A statistical analysis of the spin-split offset at particular values of $B_\parallel$ was obtained using an ensemble of traces, $\delta \cond(\gate)$, collected for a range of out-of-plane fields, 3mT$\leq$$B_\perp$$\leq$$120$mT ($\ll$$B_\parallel$).
The offset is barely visible in the raw conductance data (Fig. 2a), but can be seen more clearly by computing an autocorrelation
$ \corr(\Delta\gate)= \left\langle \delta \cond(\gate) \delta \cond(\gate+\Delta\gate) \right\rangle$.
Side-peaks in the autocorrelation at $\Delta\gate = \pm \voff$ (Fig.~2b) reflect similar features offset in the conductance traces (Fig.~2a) due to the difference in spin-up and spin-down carrier densities.

The side-peaks shifted outward with increasing $B_\parallel$ (Fig.~3a), while the height of the central peak ($\corr(0)$, the variance of the fluctuations) dropped by a factor of two (Fig.~3b).  The drop in $\corr(0)$ reflects a suppression of conductance fluctuations when the Zeeman energy, $\gfactor\bohr B$, exceeds energy broadening due to temperature or dephasing. This effect has been used to characterize spin degeneracy in other systems\cite{PhysRevB.56.15124,PhysRevLett.63.2264,PhysRevLett.86.2102}, where splitting could not be observed directly.   The suppression factor of $\sim$$2$ indicates that spin degeneracy was intact at zero field, after taking into account broken time reversal symmetry due to the small out-of-plane field.  The shift of the side-peak location, $\voff$, with $B_\parallel$ was expected from Eq.~1---a statistical confirmation of the linear spin-splitting seen in Figs.~1F and 1G.

The density of states can be extracted directly from $\voff$ at a given field using $g$=$2$ for graphene (see supplement) and the value for $\alpha$ determined from quantum Hall measurements (Eq.~1).  $\voff(\gate)$ was recorded over the full gate voltage range by computing the autocorrelation $\corr(\Delta\gate)$ within a sliding window in $\gate$ (Fig.~3c).   The resulting square-root lineshape can be compared to $\dos$ expected from graphene's dispersion relation:
$\dos(\gate) = \frac{2\sqrt{\alpha/\pi}}{\hbar \speed} \sqrt{ |\gate-V_0|}$,
with a Fermi velocity $\speed$ that
would be independent of density for an ideal Dirac band structure.

A quantitative analysis of $\voff(\gate)$ gives $\speed$$>$$1.0$$\times10^6$m/s over the range of densities reported here (Fig.~3d), consistent with measurements using a variety other techniques\cite{deacon:081406,infrared,2008NatPh...4..144M}, but considerably larger than the expected $\speed \approx 0.8$$\times$$10^6$m/s from non-interacting band structure calculations\cite{PhysRevB.72.174406}.
Enhancements in $\speed$ have been predicted due to many-body effects\cite{sarma:121406,trevisanutto:226405},
which become particularly important at low densities.  Indeed, larger values of $\speed$ are observed in Fig.~3d at lower densities, similar to a report of infrared spectroscopy measurements\cite{infrared}. An additional deviation from the ideal Dirac density of states is seen at charge neutrality (from $\gate$ $\sim$ 0--2V): rather than dropping to zero, a finite value of $\voff$ is visible (Fig.~3c, supplement).
This minimum density of states, $4\times10^{12}$ cm$^{-2}$/eV, is similar to that observed in scanning single-electron transistor measurements\cite{2008NatPh...4..144M}, attributed to broadening from disorder.

In order to observe V's like those shown in Figs.~1F and 1G, interference features must shift, but not otherwise change, as a function of in-plane field.  In other words, the primary influence of the magnetic field on the interference must be through its effect on the densities of spin-up and spin-down carriers, rather than a change in Aharonov-Bohm (AB) flux.  Fields of a Tesla or more were required to differentiate the two carrier densities, but the AB flux introduced by a few milliTesla in $B_\perp$ was sufficient to alter the observed interference pattern.   The drastically different field scales for spin and AB effects demanded precise alignment (within $0.05^\circ$) of the $B_\parallel$ axis to the plane of the graphene flake.
The alignment of $B_\parallel$ was monitored using weak localization (WL), the coherent enhancement of backscattering associated with time-reveral symmetry at $B_\perp$=0\cite{beenakker}.
Slight mis-alignment led to a shift in the location of the WL conductance dip as $B_\parallel$  was raised, and was corrected by offsetting $B_\perp$.

In addition to shifting, the WL dip also decreased in magnitude with $B_\parallel$, by a factor of two at 1T and below detectable levels above 4T (Fig.~4ab).
The complete collapse of the WL dip, over a field range where the variance of conductance fluctuations decreases only by a factor of two, implies that time-reversal symmetry is broken even by an in-plane field\cite{zumbuhl,PhysRevB.35.2844}.
The disappearance of symmetry in $g(B_\perp)\leftrightarrow g(-B_\perp)$ at finite $B_\parallel$ (Fig.~4cd) provided further evidence of time-reversal symmetry breaking\cite{zumbuhl}.  Similar phenomena have been observed in semiconductor 2D electron gases, and have been associated with finite thickness and nanometer-scale undulations in the 2DEG\cite{PhysRevB.35.2844}, allowing $B_\parallel$ to thread an AB flux through the conductor.
The $B_\parallel$-scale of WL collapse in graphene corresponds to an effective thickness of $\sim$1nm, in agreement with previous measurements of the intrinsic ripple size in graphene sheets\cite{2007Natur.446...60M,nanolettripples}.

We acknowledge V.~F'alko, D.~Goldhaber-Gordon, H.~Heersche, T.~Ihn, G.~Kamps, C.-Y.~Lo, A.~Morpurgo, and K.~Todd for helpful discussions.  Graphite provided by S.~Fain and D.~Cobden.  MBL acknowledges a PGS-D from NSERC; work funded by CIFAR, CFI, and NSERC.

\clearfig
\begin{center}
\includegraphics{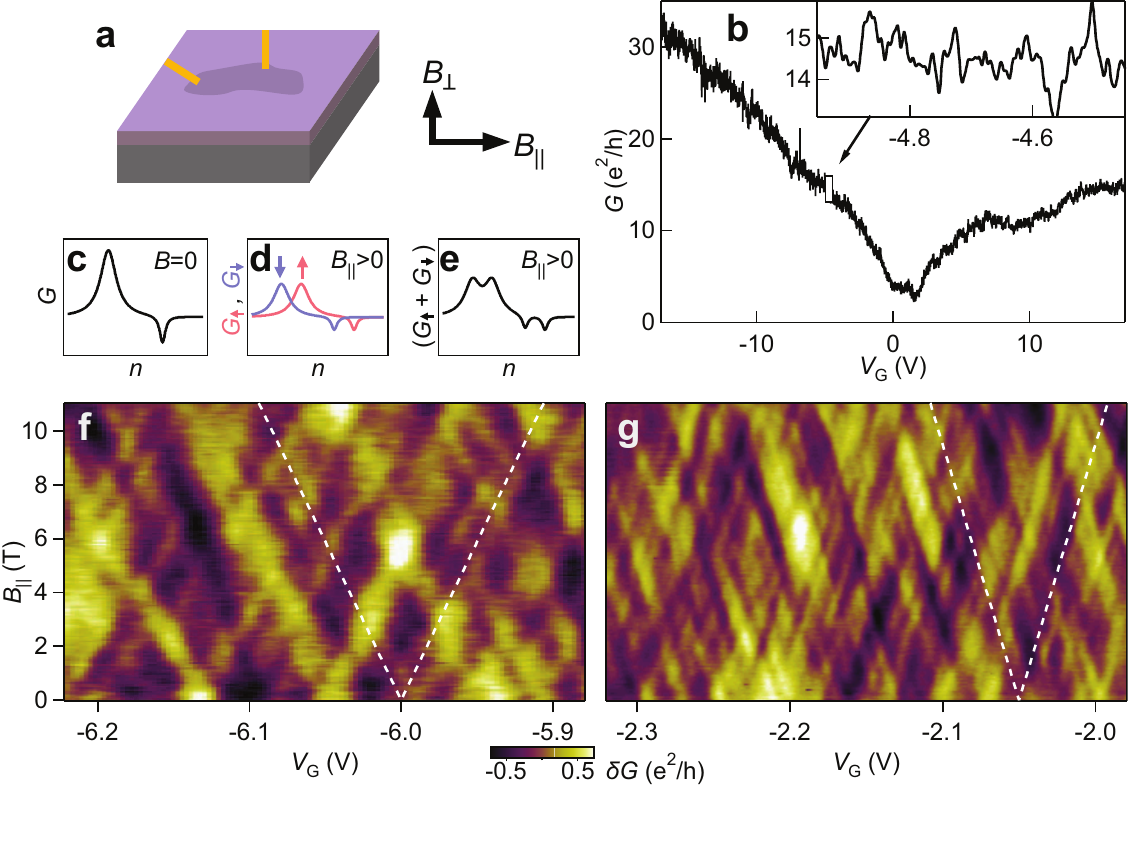}
\end{center}

\noindent {\bf Fig.~1.
Zeeman effect in graphene.}
({\bf a}) Schematic of single-layer graphene device, showing orientation of magnetic fields $B_\parallel$ and $B_\perp$.
({\bf b}) Two-terminal conductance, $\cond$, measured as a function of backgate voltage $\gate$ at 20mK, showing the Dirac conductance minimum ($B_\perp$=$B_\parallel$=20mT).
Inset: a sample of conductance fluctuations over a narrow range of gate voltage. 
~~({\bf c},{\bf d},{\bf e}) Diagrams of spin-splitting effect: a simulated conductance trace at zero field (c); two offset contributions from spin-up and spin-down at finite field (d); contributions add together in the total conductance (e), giving spin-split features.
({\bf f},{\bf g}) Two examples of conductance fluctuations spin-splitting in an in-plane field, taken over two ranges of gate voltage ($B_\perp$=80mT). For both images, a smooth background was subtracted to highlight fluctuations. The offset (indicated by dashed lines as guides to the eye) due to spin is larger around -6V compared with -2V because density of states increases with density: the dashed lines correspond to densities of states of (f) 12$\times$10$^{12}$ and (g) 7$\times$10$^{12}$cm$^{-2}$/eV.

\clearfig
\begin{center}
\includegraphics{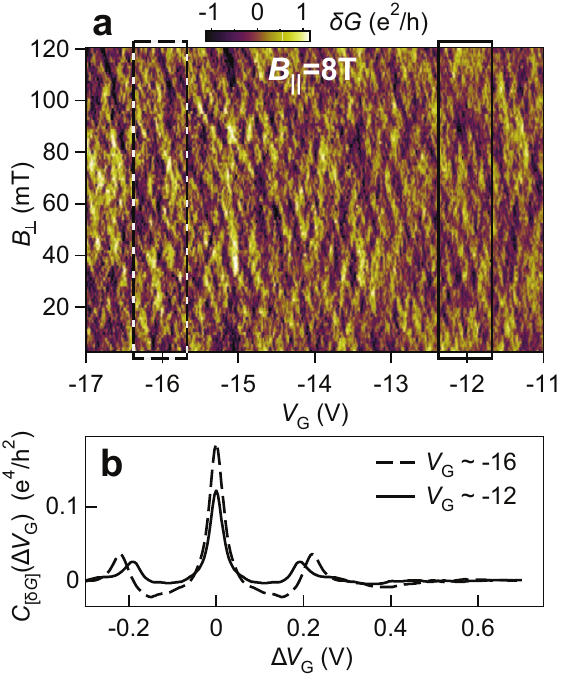}
\end{center}

\noindent {\bf Fig.~2.
Autocorrelation extraction of spin splitting.}
({\bf a}) Features are doubled in an image of conductance fluctuations at $B_\parallel$=8T.
({\bf b}) Autocorrelations of the windows indicated in (A) show that the spin-split offset (the location of the autocorrelation side-peak) changes from $\voff$=0.22V at $\gate$=-16V, to $\voff$=0.19V at $\gate$=-12V. The autocorrelation function is symmetric, so side-peaks appear at $\pm\voff$.

\clearfig
\begin{center}
\includegraphics{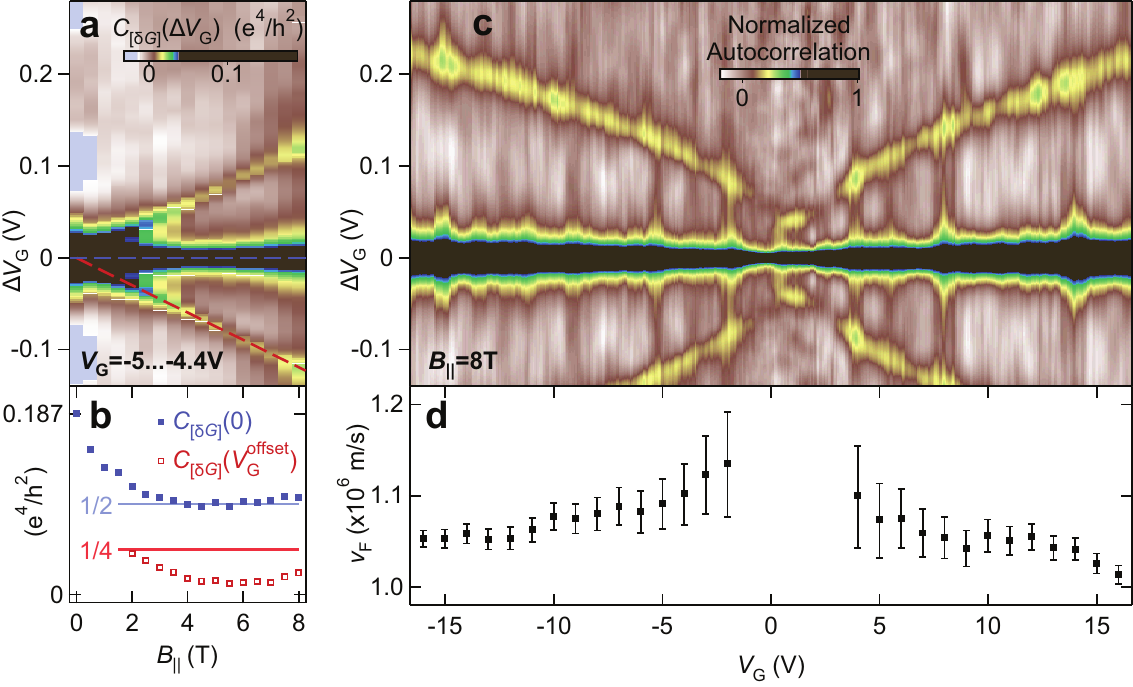}
\end{center}

\noindent {\bf Fig.~3.
Magnetic field- and gate-dependence of spin splitting.}
({\bf a}) Autocorrelations, $\corr(\Delta\gate)$, of fluctuations in the range $\gate$=-5...-4.4V, averaged over an ensemble of traces for 3mT$\leq$$B_\perp$$\leq$120mT, demonstrate the in-plane field dependence of side-peak location, $\voff$. The dashed red line is fit to $\voff$, and corresponds to a density of states 1.04$\times$10$^{13}$cm$^{-2}$/eV.
({\bf b}) Variance $\corr(0)$ (filled squares) and autocorrelation side-peak height $\corr(\voff)$ (open squares) from (a). Variance drops to the expected 50\% of its zero-field value, but the side-peaks fall below the expected 25\%.  The out-of-plane field $B_\perp$$\geq$3mT was large enough to break time reversal symmetry even at $B_\parallel=0$.
({\bf c}) Autocorrelations computed for a sliding 0.75V-wide window in $\gate$ ({\it c.f.} Fig.~2), and averaged over an ensemble of traces for 3mT$\leq$$B_\perp$$\leq$120mT, show how the density of states depends on density.
({\bf d}) The Fermi velocity $\speed$ extracted from side-peak locations, $\voff$, in (c) is enhanced near the Dirac point.
Error bars are primarily due to uncertainty in the charge neutrality point $V_0$=1.0$\pm$0.3V.

\clearfig
\begin{center}
\includegraphics{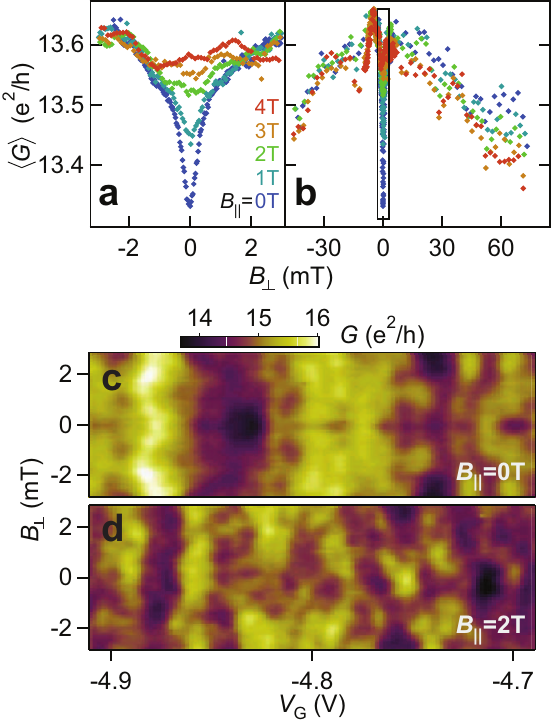}
\end{center}

\noindent {\bf Fig.~4.
Loss of weak localization and symmetry due to in-plane field.}
({\bf a}) Averaging out the conductance fluctuations over a range $\gate$=-5...-3V, a weak localization dip in conductance is present at $\vec{B}$=0 but disappears for increasing $B_\parallel$ or $B_\perp$.  This indicates a breaking of time-reversal symmetry from either component, but at very different field-scales.  Curves at different $B_\parallel$ are vertically offset to align $G(B_\perp$$>$10mT$)$ (see supplement). ({\bf b}) Negative magnetoconductance for $B_\perp$$>$10mT is unaffected by $B_\parallel$.  Box encloses range from (a).
({\bf c},{\bf d}) Conductance is symmetric in $B_\perp$ at $B_\parallel$=0 (c) but not at $B_\parallel$=2T (d).  These data provide further evidence of time-reveral symmetry breaking by the in-plane field, excluding (for example) that the disappearance of weak localization in (a) and (b) is due to decoherence.


\clearpage
{\Large\bf Supplementary information}

\textbf{[Device Fabrication]}
Graphene was deposited on SiO$_2$/Si wafers made from highly-arsenic-doped silicon with resistivity $<$ 0.007 $\Omega\cdot$cm, thermally oxidized
by the supplier to a uniform depth of 300nm$\pm$15\%.
Before graphene deposition, the oxide was thinned using a CF$_4$/O$_2$ plasma to change the colour from the initial teal to purple, $\sim265$nm,
to aid in the identification of deposited graphene. After thinning,
wirebonding pads and alignment features were laid down by standard optical lithography techniques.  The surface was cleaned with a standard SC-1/SC-2 etch immediately before graphene deposition.

\begin{figure}[ht]
\centering
\includegraphics{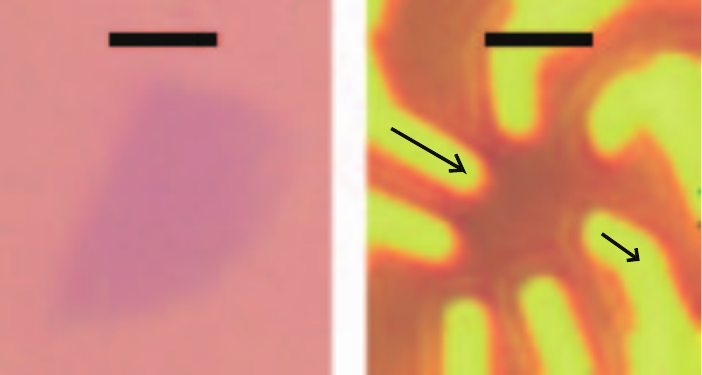}
\caption{
Optical microscope images of the graphene flake on SiO$_2$ before and after microfabrication.
Scale bars: 3$\mu$m. Arrows indicate the two contacts used in the experiment. The other wires did not make contact at low temperature.
}
\label{sfigdev}
\end{figure}

Natural graphite flakes were cleaved repeatedly by Scotch tape, then pressed lightly on the chip surface. After 10 minutes, the tape was slowly peeled off the chip, then the chip was placed in hot acetone for 2 minutes to remove nearly all the tape residue, followed by isopropanol rinse and drying.  Graphene flakes were identified by their colour in an optical microscope, where colour shifts were observed in roughly equal steps for 0,1,2,3 layers.  After identification, flakes were contacted by thermally evaporated 5nm Cr / 50nm Au metallization, patterned using an electron-beam lithography process at 20keV with a 200K/950K PMMA bilayer resist (Supp.~Fig.~\ref{sfigdev}).

\begin{figure}[ht]
\centering
\includegraphics{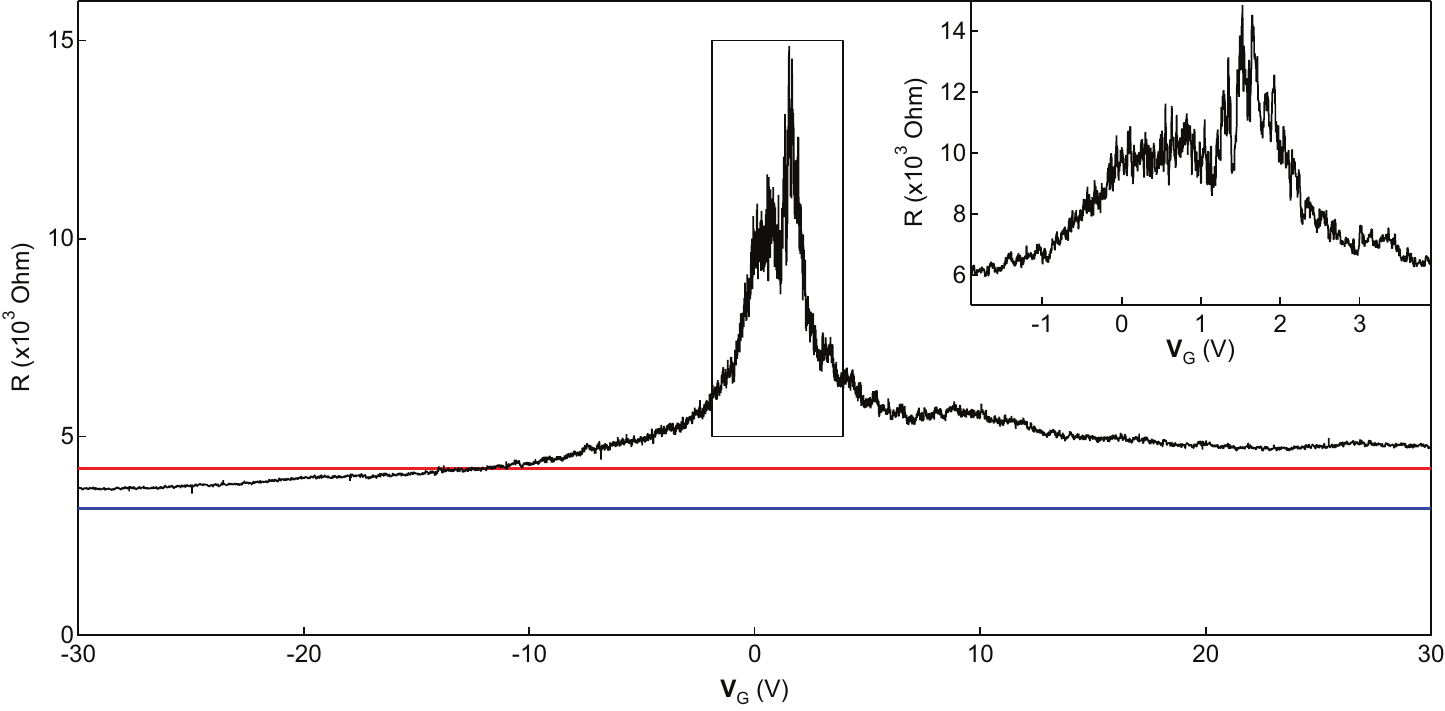}
\caption{
Dirac peak in resistance from 20mK cooldown.
The Dirac peak extrapolates to a different contact resistance for highly negative (blue, $\sim$3200$\Omega$) or highly positive gate voltage (red, $\sim$4200$\Omega$).
Inset: A slight splitting in the peak is observed, possibly due to surface contamination of the graphene.
}
\label{sfigdir}
\end{figure}

Several single-layer flakes were evaluated based on the width and offset\cite{tan:246803} of the Dirac resistance peak $R(\gate)$.  Typical widths were 5-10V, centered around a voltage often less than 5V.  These characteristics were achieved only with the SC-1/SC-2 etch step mentioned above---without this step, flakes were $\sim$50V wide and centered at $>$30V.  The flake measured in this paper had a narrow but slightly asymmetric Dirac peak, with local maxima at +0.5V and +1.5V (Supp.~Fig.~\ref{sfigdir}).  

\begin{figure}[htp]
\centering
\includegraphics{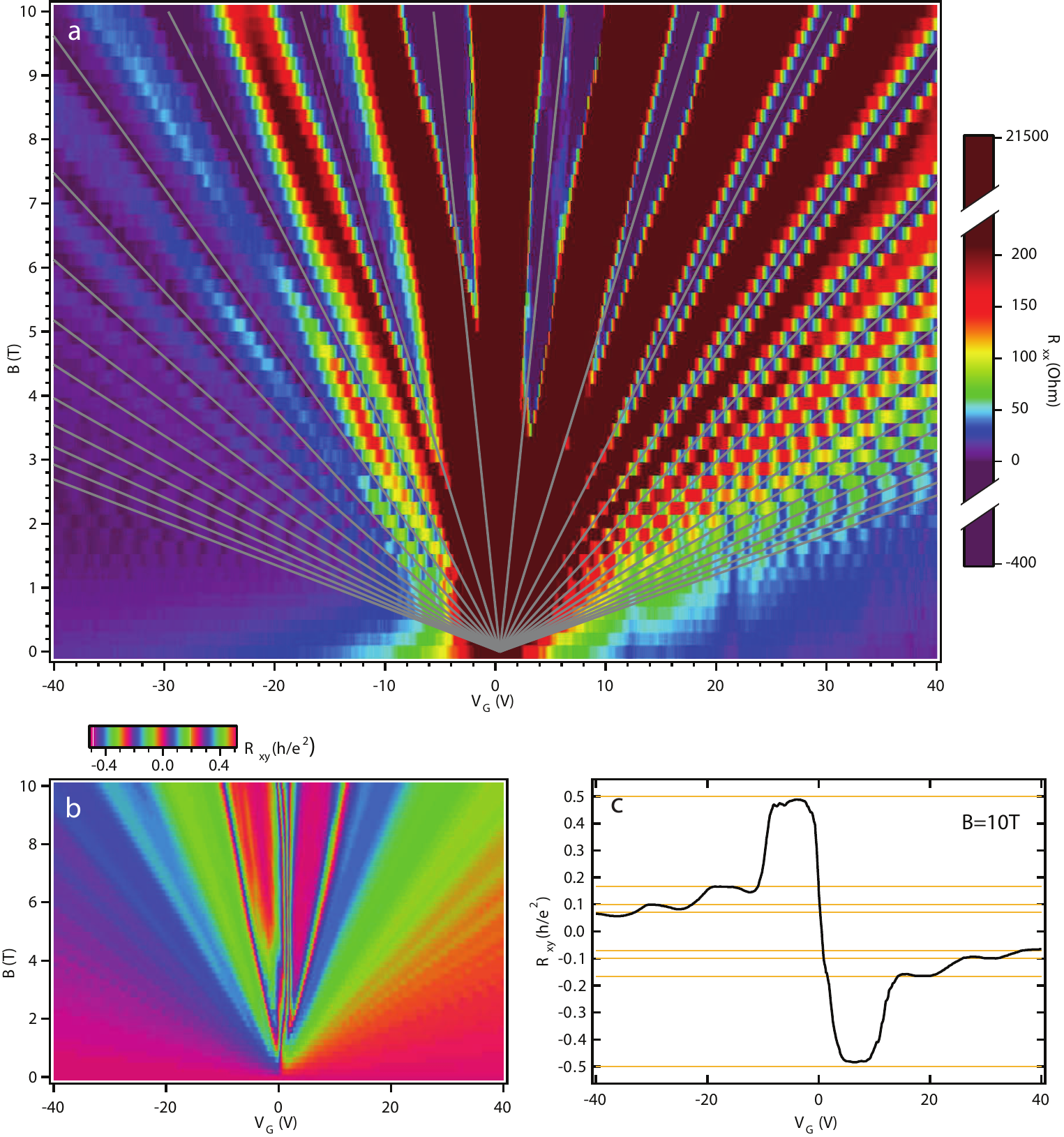}
\caption{
Quantum Hall effect.
({\bf a}) Longitudinal resistance as a function of gate and perpendicular
magnetic field. Colour scale is chosen to emphasize Rxx minima due to SdH oscillations.
The fan of lines corresponds to $\alpha$=8.05$\times$10$^{10}$ cm$^{-2}$/V and charge neutrality at +0.4 V.
({\bf b},~{\bf c}) Transverse hall resistance, showing the characteristic (n+1/2)
quantum Hall plateaux that are found only in single-layer graphene.
}
\label{sfigqh}
\end{figure}

\textbf{[4K and 20mK cooldowns]}
The flake was first measured at 4K in a quantum Hall configuration.
The characteristic $n+1/2$ quantization was observed, confirming that the flake was single layer graphene\cite{novoselov} (Supp.~Fig.~\ref{sfigqh}).  The gate dependence of Shubnikov-de-Haas oscillations allowed an accurate determination of the induced carrier density coefficient $\alpha$ (Supp.~Fig.~\ref{sfigqh}). The measured value of $\alpha$ was crucial for quantifying the density of states and Fermi velocity throughout the paper. The linear relationship $n=\alpha \gate$ is a result of the geometrical capacitance $c_{\rm g} = \alpha e = (\epsilon_{\rm SiO2}/{\rm thickness})$ between the backgate and graphene. In general, with finite densities of states an additional nonlinear `quantum capacitance' $c_{\rm q}=e^2 \dos$ is present in series with the geometrical capacitance\cite{2008arXiv0812.3927C}. Since the gate oxide was thick, the correction to total capacitance was only 2\% for the lowest density of states observed. Hence, variations of $\alpha$ were ignored in our analysis.

During the transfer from the 4K probe to the dilution refrigerator one of the contacts to the graphene was broken, preventing a 4-wire resistance measurement.  As a result, the data presented in the main paper are two-terminal measurements.

\textbf{[Conductance Measurement]}
Dilution refrigerator measurements were performed using 10nA lock-in excitation at 77.92 Hz. A contact resistance of 3200 $\Omega$ was estimated by extrapolating the Dirac peak resistance to large negative gate voltages (Supp.~Fig.~\ref{sfigdir}); this resistance was subtracted before calculating the conductance.  The Dirac peak extrapolated to 4200 $\Omega$ for positive gate voltages, perhaps due to a diode effect in the contacts. This $\gate$-dependent contact resistance does not affect our analysis, as it is insensitive to the absolute amplitude of $\delta G$ fluctuations. Mobility was estimated using the conductance data in Fig.~1b and assuming a two-square geometry, giving 4000 cm$^2$V$^{-1}$s$^{-1}$ for high carrier densities.

\begin{figure}[ht]
\centering
\includegraphics[scale=0.7]{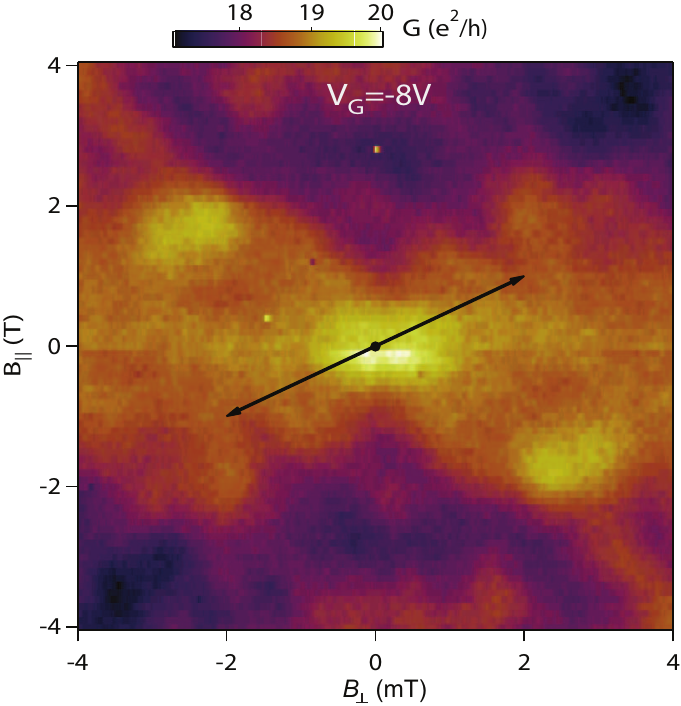}
\caption{
Symmetry in conductance under field reversal $\vec B \leftrightarrow -\vec B$, as required by reciprocity relations.
}
\label{sfigbb}
\end{figure}

\textbf{[Field orientation]}
The two-axis magnet used for dilution refrigerator measurements had a primary solenoid aligned to within $\sim0.4^\circ$ of the plane of the sample, controllable up to $B_\parallel$=12T, and a split-coil magnet providing a field up to $B_\perp$=120mT perpendicular to the sample plane.  The separate effect of the two magnetic field components can be seen in Supp.~Fig.~\ref{sfigbb}, showing symmetry in total field reversal $\vec B \leftrightarrow -\vec B$ as required by reciprocity relations\cite{buttikerprl,buttikeribm} for a two-terminal measurement.

The $B_\perp$ magnet was used to correct for the $0.4^{\circ}$ error in $B_\parallel$ alignment, measured precisely by the shift in the weak localization conductance (WL) dip and symmetry center.  Although WL could not be measured accurately for $B_\parallel>3$T, the $B_\perp$ correction above that field assumed the same mis-alignment angle as measured at lower field.  Throughout the paper, the stated $B_\perp$ is the compensated field, as experienced by the graphene.  Although the V's from Figs.~1f and 1g were sensitive to small errors in $B_\parallel$ alignment, the autocorrelations of gate voltage traces were insensitive because they were computed for a specific $B_\parallel$, and averaged over a range of fields 3mT$\leq B_\perp \leq 120$ mT.

\textbf{[Background subtraction]}
In order to extract the fluctuations $\delta G = G - \langle G \rangle$ from the conductance data $G$, it was necessary first to evaluate $\langle G \rangle$, the conductance background.  Background subtraction was an important step in the calculation of autocorrelation functions, and had to be done carefully to avoid introducing artifacts.   Two different background extraction protocols were used, for data sets covering very different parameter ranges,  and are described below.    In all cases, the details of background subtraction influenced the $\delta G$ variance and clarity of the autocorrelations but did not affect the position of the sidepeaks, that is, the background did not affect the extracted $\voff$.

\begin{figure}[ht]
\centering
\includegraphics{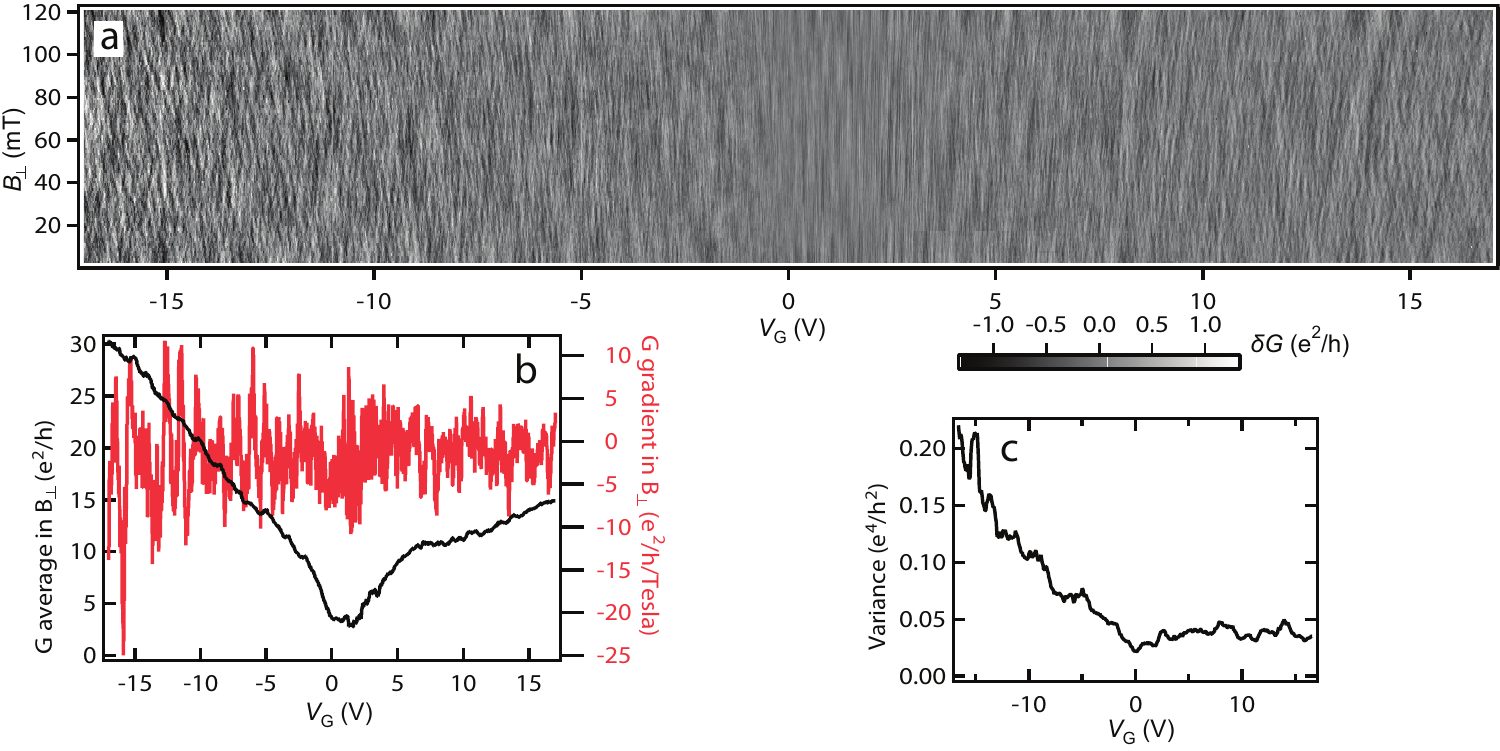}
\caption{
({\bf a}) Extracted fluctuations used in Figs.~2 and 3C. Electronic version may be zoomed to show full details.
({\bf b}) An offset and gradient in $B_\perp$ for each value of $\gate$ were subtracted from the raw conductance data in order to obtain (a).
({\bf c}) Variance of a, calculated within a sliding window of 0.75V.}
\label{sfigbgh}
\end{figure}

The data in Figs.~2 and 3c were at $B_\parallel=8$T, but covered a wide range of $\gate$. (Supp.~Fig.~\ref{sfigbgh}).  For these data sets, a linear function $\langle G\rangle(B_\perp)$ was fit at each value of $\gate$ and subtracted off.  Since the range of $B_\perp$ encompasses many conductance fluctuations, this procedure does not remove the primary signal, but it does remove larger gate- and field-scale fluctuations that would otherwise overwhelm the autocorrelation.

For Fig.~3a, which covered only a narrow range of densities but  a wide range of in-plane fields, the average effect on conductance of each of the three control parameters---$\langle G \rangle (\gate)$, $\langle G \rangle (B_\perp)$, and $\langle G \rangle (B_\parallel)$---was calculated by averaging fluctuations over the other two parameters.  For example, $\langle G \rangle (B_\parallel)$ was calculated by averaging over both $\gate$ and $B_\perp$, denoted $\langle G \rangle_{\gate,B_\perp} (B_\parallel)$. Then the three-dimensional function  $\langle G \rangle(\gate,B_\perp,B_\parallel)$ was calculated by adding the three effects together. The subtracted background was
$ \langle G \rangle(\gate,B_\perp,B_\parallel) =
\langle G \rangle_{B_\perp,B_\parallel} (\gate) +
\langle G \rangle_{B_\parallel,\gate} (B_\perp) +
\langle G \rangle_{\gate, B_\perp} (B_\parallel) - 
2\langle G \rangle_{\gate, B_\perp,B_\parallel} $. 

\textbf{[Graphene g-factor]}
Graphene is expected to have a $g$-factor close to that of the free electron, but to our knowledge graphene's $g$-factor has been measured only in the quantum Hall regime\cite{graphenehigh}, where Landau quantization can lead to significant modifications\cite{exchgfactor}. On the other hand, extensive spin resonance measurements in graphite have found $\gfactor_{\parallel}$=2.0029 for in-plane fields\cite{Stankowski2000489,PhysRev.118.647,PhysRevB.44.11845}.  We therefore approximate $\gfactor$=2 throughout the paper.

\begin{figure}[htp]
\centering
\includegraphics[scale=0.9]{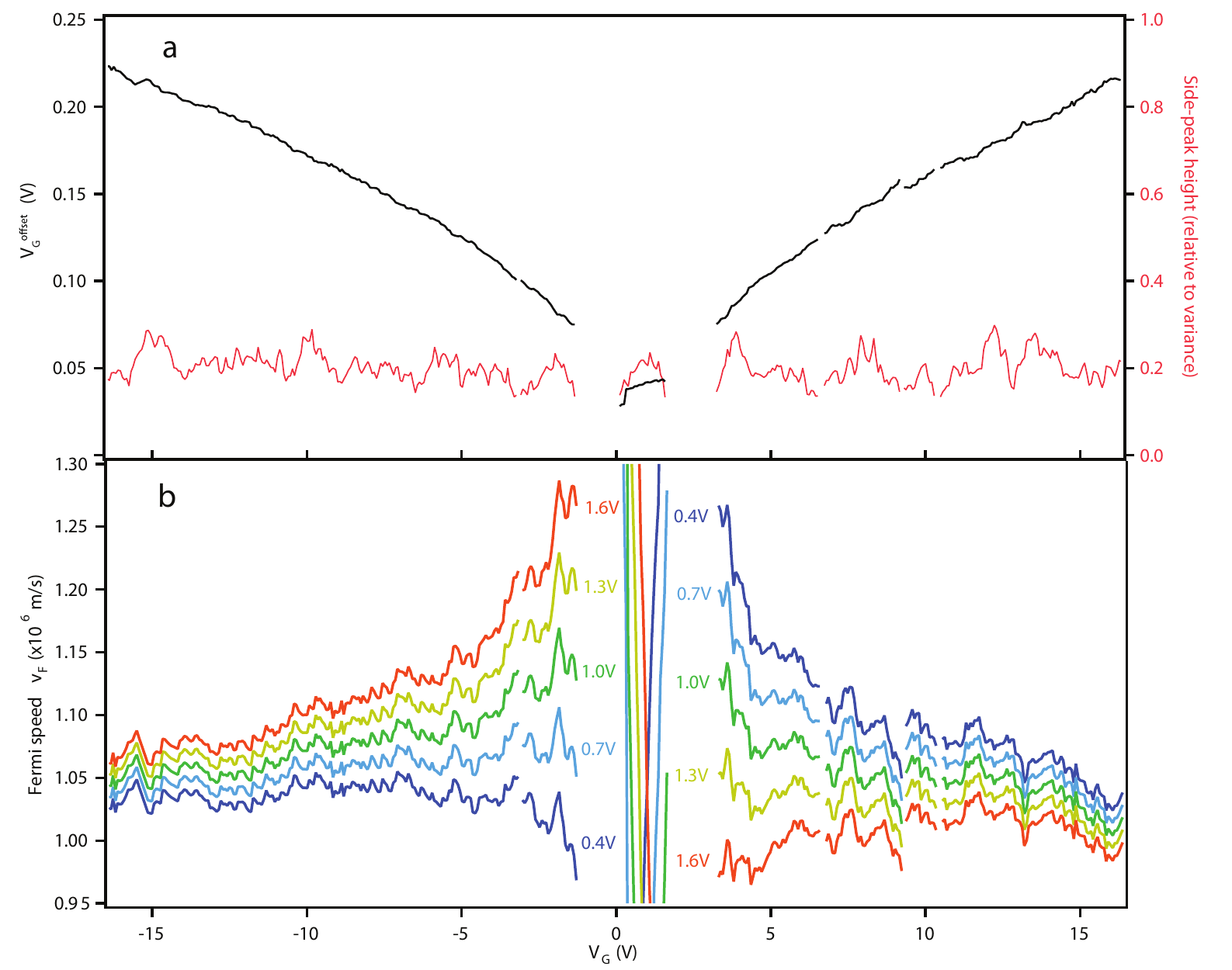}
\caption{
({\bf a}) Offsets $\voff$ extracted from Fig.~3c autocorrelations, along with autocorrelation side-peak heights relative to the variance. Expected side-peak height is 50\% for ideal spin-splitting.
({\bf b}) Fermi speeds $\speed$ calculated from $\voff$. Determination of $\speed$ depends on the charge neutrality point $V_0$. Several possible values of $V_0$ are indicated with different colors.
}
\label{sfigext}
\end{figure}

\textbf{[Density of states and Fermi velocity]}
The density of states for any 2D isotropic dispersion relation $\epsilon(\vec k) = \epsilon\left(\sqrt{k_x^2+k_y^2}\right)$ with degeneracy $D$ can be written as $\dos=\frac{{\rm d}n}{{\rm d}k}/\frac{{\rm d}\epsilon}{{\rm d}k}=\sqrt{Dn/\pi}/\frac{{\rm d}\epsilon}{{\rm d}k}$.  Defining Fermi velocity as $\speed \equiv \frac{1}{\hbar}\frac{{\rm d}\epsilon}{{\rm d}k}$, this gives
$$\speed = \frac{\sqrt{D/\pi}}{\hbar \dos} \sqrt{n}$$
with graphene's $D=2\times 2$ from valley and spin degeneracy. The Fermi velocities reported in Fig.~3d were calculated using:
\begin{itemize}
\item the density of states $\dos = \frac{\alpha}{g\bohr B}\voff $ extracted from the spin-splitting offset $\voff$ at $B_\parallel=8$T (Fig.~3c, Supp.~Fig.~\ref{sfigext}), and
\item the density $n=\alpha(\gate-V_0)$ determined by the gate voltage $\gate$ away from charge neutrality $V_0$.

\end{itemize}

Uncertainty in $V_0$ was the primary source of the uncertainty in $\speed$ in Fig.~3d. As seen in Supp.~Fig.~\ref{sfigext}, a more positive value for $V_0$ implies a reduced $\speed$ for electrons  (enhanced for holes); a more negative value does the opposite.
The value used in the paper, $V_0=1.0\pm0.3$V, was chosen to reflect the position, width, and asymmetry of the Dirac peak (Supp.~Fig.~\ref{sfigdir}). Beyond this range, the extracted values for $\speed$ were highly asymmetric for electrons and holes at low density.

\begin{figure}[htp]
\centering
\includegraphics[scale=0.9]{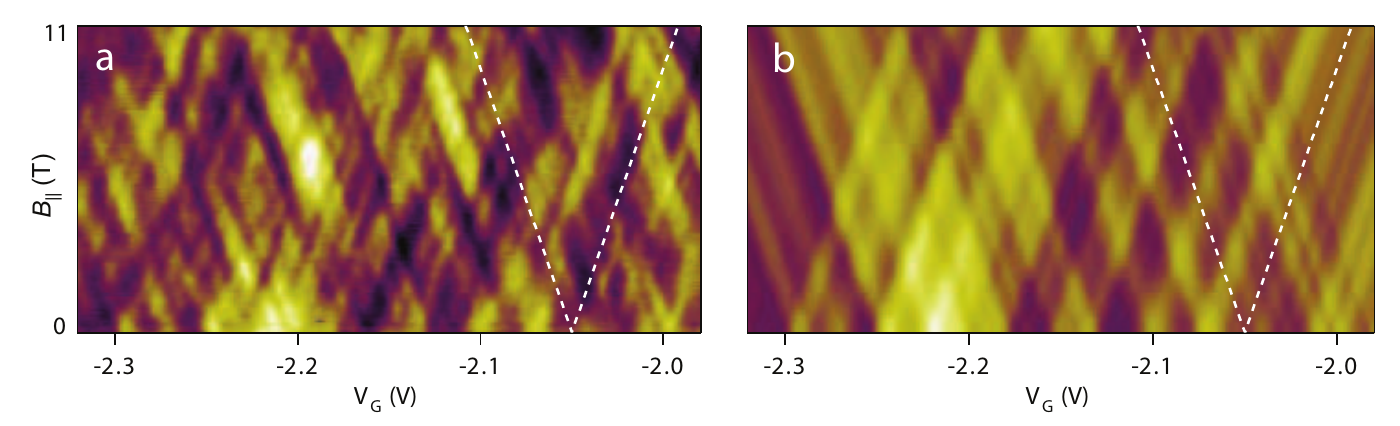}
\caption{
({\bf a}) Duplicated Fig.~1g from the paper. Spin-splitting offset increases with magnetic field.
({\bf b}) Idealized extrapolation upwards from the $B_\parallel$=0 data, showing expected appearance of the data if fluctuations split into two identical spin-split copies without otherwise changing.
}
\label{sfigsplit}
\end{figure}

\textbf{[Deviations from ideal spin splitting]}
Some experimental observations point to a more complicated spin-splitting behavior than simple picture of two offset densities described in the text, indicating that interference for spin-up and spin-down carriers is somewhat different.  First, the V's from Fig.~1f and 1g in the main text contain more features than would be expected simply from offset densities.  For comparison, the data from Fig.~1g is replotted next to a simulation of ``ideal'' splitting based on the $B_\parallel=0$ trace from Fig.~1g (Supp.~Fig.~\ref{sfigsplit}).  Some features move either left or right in gate voltage with $B_\parallel$ but do not split, that is, some left or right moving features do not appear to have an oppositely-moving counterpart.  One interpretation of this observation would be that the conductance fluctuations were partially spin-resolved even at zero field, but further work is required to explore this possibility.

Second, in Supp.~Fig.~\ref{sfigext} the side-peak amplitude in the autocorrelations is consistently smaller compared to the central peak than the 50\% that would be expected for two offset copies of random fluctuations (as in Supp.~Fig.~\ref{sfigsplit}).  One explanation would be that a shift in gate voltage changes not only the density but also the details of the scattering.  One might expect from this that the relative height of the side-peak would fall off with field, as $\voff$ gets larger and larger and the offset features appear farther and farther from each other in gate voltage.  In fact, the opposite behavior is seen in the data: the side-peak is higher at very high fields in Fig.~3b.  As mentioned above, slight mis-alignment in $B_\parallel$ would not explain the too-small side-peaks, which are taken from autocorrelations at a fixed value of $B_\parallel$.

\begin{figure}[ht]
\centering
\includegraphics[scale=0.75]{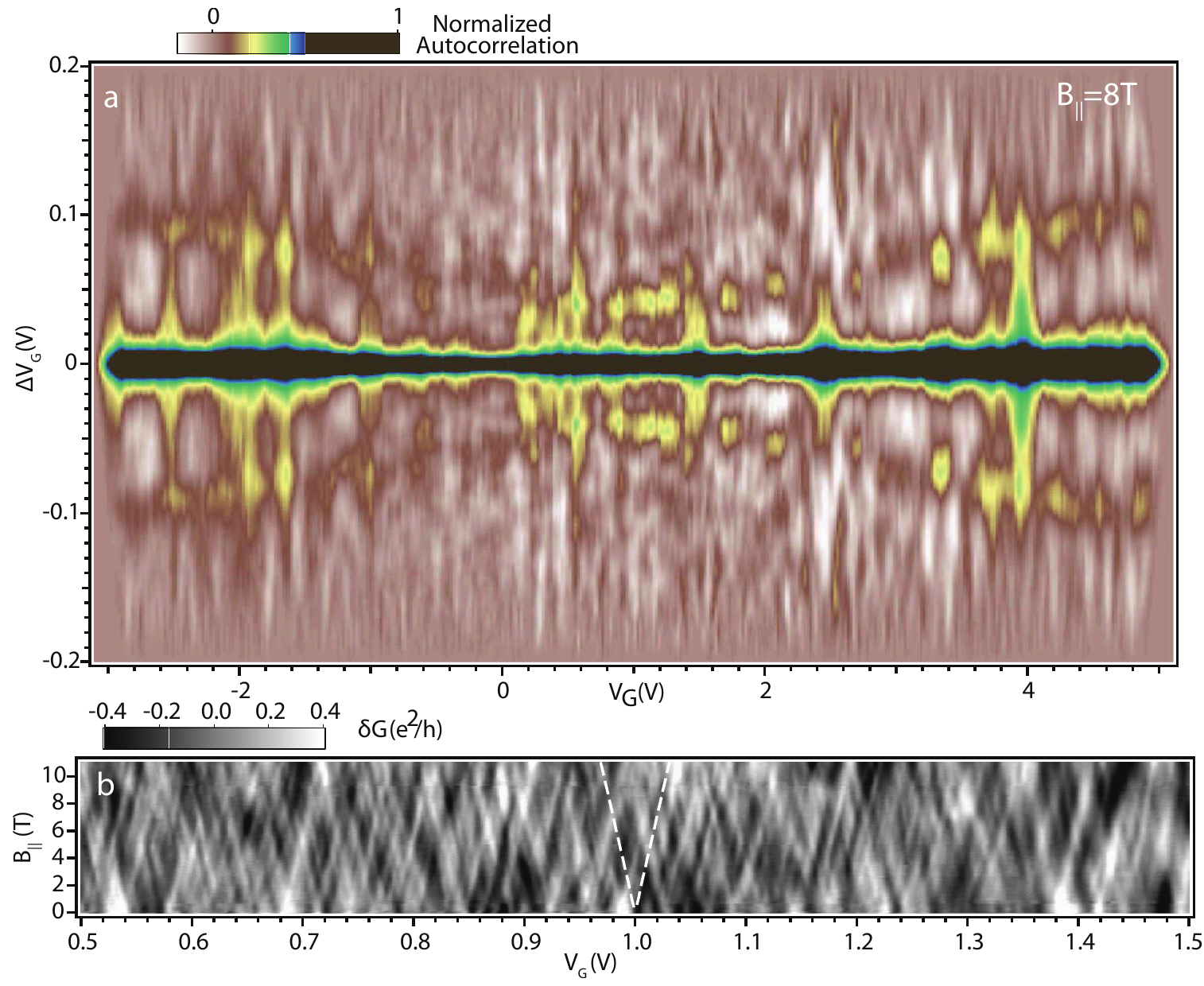}
\caption{
({\bf a})
Autocorrelations for a sliding window (similar to Fig.~3c), with a window width of 0.2 volts, and focusing on the region around the Dirac point.  A side-peak near $\voff\sim0.04$V is visible throughout the region near charge neutrality ($\gate$=0.5-1.5V), corresponding to a density of states $4\times10^{12}$cm$^{-2}$/eV.
({\bf b})  Spin splitting of features with $B_\parallel$ near charge neutrality. The dashed lines correspond to the density of states extracted from (a).  The V's in the raw data appear to follow this slope, confirming the constant density of states near charge neutrality.}
\label{sfigdpd}
\end{figure}

\textbf{[Finite density of states at charge neutrality]}
The spin-splitting offset does not drop to zero at the charge neutrality point as expected from a linear dispersion relation.  Instead, a closer examination of the $\gate=-1\ldots 3$V region from Fig.~3c (Supp.~Fig.~\ref{sfigdpd}) reveals a density of states $\dos=3.5\pm0.4\times10^9$ cm$^{-2}$/meV that is roughly independent of gate voltage. This consistent density of states is seen in both the correlation function and the `V's in the raw data. In scanning single-electron transistor measurements\cite{2008NatPh...4..144M}, the inverse compressibility (reciprocal of the density of states) was observed to reach a maximum value of about 3 $\times 10^{-10}$ cm$^2$\,meV at the Dirac point, in agreement with our measurement.

\textbf{[Ripples]}
An order-of-magnitude estimate of the ripple size can be made by comparing the in-plane and out-of-plane magnetic field scales required to suppress weak localization.
As seen in Fig. 4, the weak localization dip was supressed by 50\% for $B_\perp$ of about 0.5 mT, corresponding to a $\sqrt{(h/e)/0.5\rm{mT}} = 3\mu$m lateral phase-coherence length. In contrast, an in-plane field of approximately $B_\parallel \approx 1.5$T was required to reach 50\% suppression, indicating that the effective vertical spread of backscattering trajectories (which have 3$\mu$m lateral length) is $(h/e)/(1.5{\rm T}\times 3\mu{\rm m})=1$nm.

\begin{figure}[ht]
\centering
\includegraphics[scale=0.9]{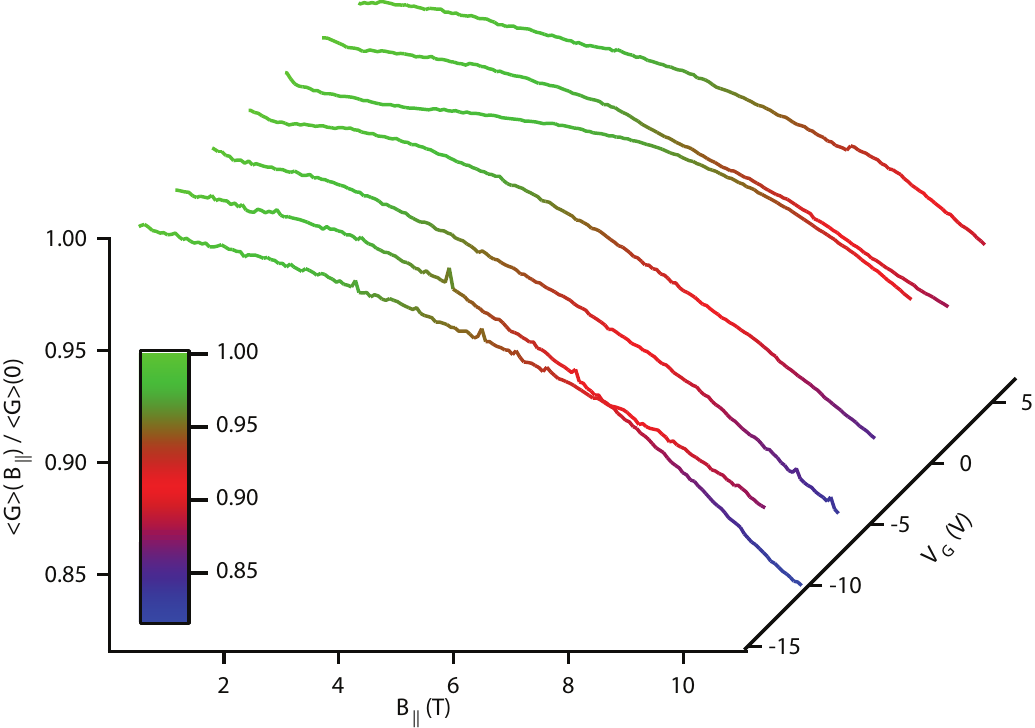}
\caption{
Negative magnetoconductance due to the in-plane field $B_\parallel$, expressed as a fraction of the zero-field conductance. The curves here were averaged in $\gate$ over 3V-wide windows centered at (from bottom to top) -13V,-10V,-7V,-4V,-1V,2V,5V.  ($B_\perp$=80mT)
}
\label{sfigbp}
\end{figure}

\textbf{[Negative magnetoconductance from in-plane field]}
A significant decrease in the conductance for larger $B_\parallel$ was observed over the full range of gate voltage, for any $B_\perp$ (Supp.~Fig.~\ref{sfigbp}). The fractional change was -10\% to -20\% for $B_\parallel$ up to 10T. An in-plane magnetoconductance has been predicted due to reduced screening in a spin-polarized system \cite{hwang-2008} but the observed effect was three orders of magnitude larger than predicted.  The origins of this in-plane magnetoconductance are not yet understood.

\clearpage
\par

\renewcommand{\figurename}{\small {\scshape Supplementary Figure}}

\clearpage

\par

\end{document}